\documentclass[aps,
               pra,
               showpacs,
               amssymb,
               nofootinbib,
               superscriptaddress,
               twocolumn,
               longbibliography]{revtex4-2}
\usepackage[ansinew]{inputenc}
\usepackage{bbold}
\usepackage{bbm}
\usepackage{bm}
\usepackage{amsbsy}
\usepackage{amsthm}
\usepackage{amssymb}
\usepackage{amsfonts}
\usepackage{amsmath, mathtools}
\usepackage{dsfont}
\usepackage{graphicx}
\usepackage{epsfig}
\usepackage{epstopdf}
\usepackage{dsfont}
\usepackage{mathrsfs} 
\usepackage{multibib}
\usepackage[dvipsnames]{xcolor}
\usepackage[colorlinks]{hyperref}
\makeatletter
\newcommand\org@hypertarget{}
\let\org@hypertarget\hypertarget
\renewcommand\hypertarget[2]{%
  \Hy@raisedlink{\org@hypertarget{#1}{}}#2%
  }
\makeatother
\usepackage[figure,table]{hypcap}
\usepackage{MnSymbol}
\usepackage{enumerate}
\usepackage{float}
\usepackage{comment}
\usepackage{braket}

\usepackage{physics}
\usepackage{subfigure}
\usepackage[normalem]{ulem}

\hypersetup{
	bookmarksnumbered,
	pdfstartview={FitH},
	citecolor={darkgreen},
	linkcolor={darkred},
	urlcolor={darkblue},
	pdfpagemode={UseOutlines}}
\definecolor{darkgreen}{RGB}{50,190,50}
\definecolor{darkblue}{RGB}{0,0,190}
\definecolor{darkred}{RGB}{238,0,0}
\usepackage{soul}

\newcommand{\brakket}[3]{\ensuremath{\langle{#1}|{#2}|{#3}\rangle}}

\newcommand{\subtiny}[3]{\ensuremath{_{\hspace{#1 pt}\protect\raisebox{#2 pt}{\tiny{$ #3$}}}}}
\newcommand{\suptiny}[3]{\ensuremath{^{\hspace{#1 pt}\protect\raisebox{#2 pt}{\tiny{$ #3$}}}}}

\newcommand{\djj}{d\kern-0.4em\char"16\kern-0.1em}

\makeatletter
\renewcommand{\p@subsection}{}
\renewcommand{\p@subsubsection}{}
\makeatother

\usepackage{orcidlink}

\definecolor{somegreen}{RGB}{25,150,25}

\begin{document}
\title{On the existence of fully inseparable biseparable Gaussian states}

\author{Olga Leskovjanov{\'a}\,\orcidlink{0000-0002-3638-3627}}
\email{ola.leskovjanova@gmail.com}
\affiliation{Department of Optics, Palack\'y University, 17. listopadu 12, 771~46 Olomouc, Czech Republic}
\thanks{O. L. and K. B. contributed equally.}
\author{Kl{\'a}ra Baksov{\'a}\,\orcidlink{0009-0009-8944-6044}}
\email{klara.baksova@tuwien.ac.at}
\affiliation{Technische Universit{\"a}t Wien, Atominstitut \& Vienna Center for Quantum Science and Technology (VCQ),  Stadionallee 2, 1020 Vienna, Austria}
\author{Jan Provazn{\'i}k\,\orcidlink{0000-0002-5646-6964}}
\email{provaznik@optics.upol.cz}
\affiliation{Department of Optics, Palack\'y University, 17. listopadu 12, 771~46 Olomouc, Czech Republic}
\author{Ladislav Mi\v{s}ta, Jr.\,\orcidlink{0000-0002-1979-7617}}
\email{mista@optics.upol.cz}
\affiliation{Department of Optics, Palack\'y University, 17. listopadu 12, 771~46 Olomouc, Czech Republic}
\author{Nicolai Friis\,\orcidlink{0000-0003-1950-8640}}
\email{nicolai.friis@tuwien.ac.at}
\affiliation{Technische Universit{\"a}t Wien, Atominstitut \& Vienna Center for Quantum Science and Technology (VCQ),  Stadionallee 2, 1020 Vienna, Austria}

\begin{abstract}
    Genuine multipartite entanglement and full inseparability are two inequivalent quantum resources. Even though all genuinely multipartite entangled states are also fully inseparable, not all fully inseparable states are genuinely multipartite entangled. There exist fully inseparable states that can be prepared as convex mixtures of states separable with respect to different bipartite splits. Here, we are interested in examples of Gaussian states that possess this type of entanglement, so-called fully inseparable biseparable states. We show for several archetypical families of multimode Gaussian states that fully inseparable biseparable candidate states are actually genuinely multipartite entangled. Using projections to finite-dimensional subspaces and fully decomposable witnesses, we observe a shrinking of the regions of potentially fully inseparable biseparable Gaussian states with growing dimension of the projection subspaces. We therefore conjecture that all fully inseparable Gaussian states are genuinely multipartite entangled.
\end{abstract}

\maketitle

\section{Introduction}\label{sec:Intro}

{\noindent}Multipartite entanglement represents a central resource for quantum information processing that is tied to applications in quantum networks~\cite{BaeumlAzuma2017}, quantum key distribution~\cite{EppingKampermannMacchiavelloBruss2017, PivoluskaHuberMalik2018}, and conference-key agreement~\cite{RibeiroMurtaWehner2018}, as well as to quantum-computational speed-ups~\cite{JozsaLinden2003}, quantum algorithms~\cite{BrussMacchiavello2011}, quantum error correction~\cite{Scott2004}, measurement-based quantum computation~\cite{RaussendorfBriegel2001, BriegelRaussendorf2001}, to name but a few examples. Much attention has therefore been devoted to the study of the structure~\cite{HuberDeVicente2013,DeVicenteSpeeKraus2013,DeVicenteSpeeSauerweinKraus2017,SlowikHebenstreitKrausSawicki2020}, transformation~\cite{SpeeDeVicenteSauerweinKraus2017},  distribution~\cite{MorelliSauerweinSkotiniotisFriis2022,SpeeKraft2024,LiSpeeHebenstreitdeVicenteKraus2024}, and detection~\cite{TothGuehne2005a, GuehneToth2009, GabrielHiesmayrHuber2010, FriisMartyEtal2018, FriisVitaglianoMalikHuber2019, LiDaiMunozAriasReuerHuberFriis2026} of multipartite entanglement in finite-dimensional systems. 
Meanwhile, studies of multipartite entanglement in infinite-dimensional continuous-variable (CV) systems, especially for the paradigmatic class of Gaussian states (for reviews, see, e.g.,~\cite{BraunsteinVanLoock2005, WeedbrookPirandolaGarciaPatronCerfRalphShapiroLloyd2012, AdessoRagyLee2014}), have been much more focused~\cite{GiedkeKrausLewensteinCirac2001, AdessoSerafiniIlluminati2006, HiroshimaAdessoIlluminati2007}.

In this paper, we systematically examine multipartite entanglement in three-mode Gaussian states. 
Previous studies~\cite{AdessoSerafiniIlluminati2006} have focused on the characterization of states into sets of such states that are either separable with respect to one or more partitions of the system into groups of two or three modes{\textemdash}which we refer to as partition-separable states{\textemdash}or fully inseparable states that are entangled with respect to all partitions and hence multipartite entangled. 
Here, we are concerned with a more fine-grained separation of the latter set of fully inseparable states into \emph{fully inseparable biseparable} (FIB) states, which are entangled with respect to every bipartition but can nonetheless be written as convex combinations of states that are separable with respect to different partitions, and \emph{genuine multipartite entanglement} (GME)~\cite{GuehneToth2009,TehReid2014}. 

In particular, we focus on Gaussian states whose covariance matrix (CM) matches that of non-Gaussian states, which are (by construction) biseparable. 
This is motivated by the observation~\cite{BaksovaLeskovjanovaMistaAgudeloFriis2025} that such Gaussian states can feature GME, despite being fully determined by their CM, and that there are (non-Gaussian) biseparable states with the same CM. 
As such, the GME of the Gaussian states in question cannot be detected by what we call the CM biseparability criterion, first introduced in~\cite{HyllusEisert2006}, and indeed, not by any other biseparability criterion that can be phrased as a test of the first and second moments only for arbitrary states, as we will elaborate on in the following.

More specifically, we perform a systematic investigation across several families of three-mode Gaussian states that are fully inseparable, yet satisfy the CM biseparability criterion: For every biseparable state with a given CM one may find a second matrix that decomposes into a convex sum of CMs corresponding to Gaussian states separable with respect to different partitions, such that the difference of the original and the second matrix is positive semidefinite.
Specifically, we consider states whose CM arises from mixing the CMs of partition-separable three-mode states composed of a two-mode squeezed vacuum (TMSV) state shared by two entangled parties and a single-mode squeezed, coherent, or thermal state of the third party, and states whose CM is obtained by mixing the CMs of pure Gaussian genuinely multipartite entangled states analogous to Greenberger{\textendash}Horne{\textendash}Zeilinger (GHZ) states~\cite{VanLoockBraunstein2000} (so-called GHZ-like states) with vacuum noise.

We employ different biseparability criteria to test for GME among these states and find that, among the tested criteria, the only successful approach for detecting GME in these Gaussian states is based on obtaining Fock-state density-matrix elements from their quasiprobability distributions. This allows us to use a few of the established GME criteria that either depend only on a small subset of density-matrix elements, such as~\cite{GabrielHiesmayrHuber2010}, or can be applied on a finite-dimensional state~\cite{JungnitschMoroderGuehne2011a} obtained from local projections to finite-dimensional subspaces of the three-mode Hilbert space. We also compare the efficiency of these different approaches.

For all tested families, we find that Gaussian states whose CMs correspond to FIB non-Gaussian states are genuinely multipartite entangled for a certain non-trivial range of parameters. Moreover, this range increases with the dimension of the finite-dimensional subspaces used in the local projections. Taken together, the absence of any explicit counterexample (or strategy to construct one) and the observed behavior of the GME detection with increasing projection dimension suggest that Gaussian FIB states may not exist. Although our results do not themselves exclude their existence, and this question therefore remains open, they strongly support our conjecture that FIB Gaussian states do not exist. If true, the conjecture would imply that full inseparability and GME are equivalent for Gaussian states, which would significantly simplify the problem of detecting GME in such states. At the same time, it would follow that the recently discovered phenomenon of multi-copy GME activation~\cite{YamasakiMorelliMiethlingerBavarescoFriisHuber2022}{\textemdash}which has been shown to be generically possible in all finite-dimensional~\cite{PalazuelosDeVicente2022} and infinite-dimensional~\cite{BaksovaLeskovjanovaMistaAgudeloFriis2025} systems and which has also been experimentally validated~\cite{StarekGollerthanLeskovjanovaMethTirlerFriisRingbauerMista2026,ZhangGuehnePan2025}{\textemdash}would not exist for Gaussian states.

The paper is organized as follows. In Sec.~\ref{subsec:gaussianstates}, we give a brief introduction to the formalism of 
\textcolor{somegreen}{CV} systems and Gaussian states. In Sec.~\ref{subsec:entanglement} we explain the basics of multipartite entanglement theory. In Sec.~\ref{sec:results}, we present in detail the results of GME detection in the investigated families of Gaussian states. Finally, Sec.~\ref{sec:Conclussion} presents the discussion and conclusion.

\section{Gaussian states}\label{subsec:gaussianstates}

{\noindent}Gaussian states are quantum states of one or more CV systems, and we will refer to these systems as modes. In this paper, we will restrict ourselves to three-mode states with the modes labeled $A$, $B$, and $C$, respectively. To each mode~$j$ are associated quadrature operators $x_j$ and $p_j$ satisfying $\left[x_j,p_k\right]=i\delta_{j,k}$. The quadrature operators can be arranged in a vector $\xi=\left(x_A, x_B, x_C, p_A, p_B, p_C\right)^\top$, and we can rewrite the commutation relation in terms of the vector $\xi$ as $\left[\xi_j,\xi_k\right]=i\left(\Omega\right)_{jk}$ where $\Omega$ is the symplectic form given by
\begin{equation}\label{eq:omega3}
    \Omega=\left(\begin{array}{cc}
        \mathbb{0} & \mathbb{1} \\
        -\mathbb{1} & \mathbb{0}
    \end{array}\right),
\end{equation}
where $\mathbb{1}$ and $\mathbb{0}$ are the $3\times3$ identity and the zero matrix, respectively.

Every $N$-mode state $\rho$ is fully characterized by the Wigner function \cite{Wigner1932}
\begin{align}\label{eq:wignerGen}
    W(\mathbf{x},\mathbf{p})\left[\rho\right]=\left(2\pi\right)^{-N}\int\!\!\dd^{N}\!\mathbf{x'}\ e^{i\mathbf{x'}\cdot\mathbf{p}}
    \bigl\langle\mathbf{x}-\tfrac{\mathbf{x'}}{2}\bigr|\hspace*{1pt}\rho\hspace*{1pt}\bigl|\mathbf{x}+\tfrac{\mathbf{x'}}{2}\bigr\rangle,
\end{align}
with $\bigl|\mathbf{x}\pm\tfrac{\mathbf{x'}}{2}\bigr\rangle=\bigotimes_{j=1}^N \bigl|x_j\pm\tfrac{x'_j}{2}\bigr\rangle$, where $\ket{x_j}$ are eigenstates of $x_j$.
Gaussian states are defined as CV states with a Gaussian Wigner function
\begin{equation}\label{eq:wignerGauss}
    W_G(\mathbf{r})=\frac{e^{-(\mathbf{r}-\mathbf{d})^\top\gamma^{-1}(\mathbf{r}-\mathbf{d})}}{\pi^N\sqrt{\det \gamma}},
\end{equation}
where $\mathbf{r}$ is a real vector of coordinates in phase space, and the vector $\mathbf{d}$ is the vector of first moments with elements $d_j=\expval{\xi_j}=\Tr(\rho\, \xi_j)$. The matrix $\gamma$ is a $6\times 6$ real symmetric matrix collecting the second moments of the state and is known as the covariance matrix (CM), with elements given by $\gamma_{jk}=\expval{\acomm{\Delta \xi_j}{\Delta \xi_k}}$, where $\acomm{X}{Y}=XY+YX$ and $\Delta X=X-\expval{X}$.

Every quantum state has to satisfy the uncertainty principle, which translates to the following condition for the CM~\cite{SimonMukundaDutta1994, Simon2000}
\begin{equation}\label{eq:uncertainty}
    \gamma+i\Omega\geq0.
\end{equation}
Every real symmetric matrix satisfying $\gamma>0$ and $(\ref{eq:uncertainty})$ defines some Gaussian state up to displacements of the first moments.

All states with a Wigner function that is not of the form of Eq.~(\ref{eq:wignerGauss}) are called non-Gaussian states. One can see that any Gaussian state is therefore fully determined by its first and second moments. Moreover, first moments can be set to zero by local unitary displacements that do not affect the entanglement of the state and also leave the CM invariant. Thus, for the purpose of studying entanglement, the first moments can be assumed to be zero, and the CM alone is sufficient for the analysis of entanglement in Gaussian states.

\section{Genuine multipartite entanglement}\label{subsec:entanglement}

\begin{figure}[t]
    \centering
    \includegraphics[width=0.7\linewidth]{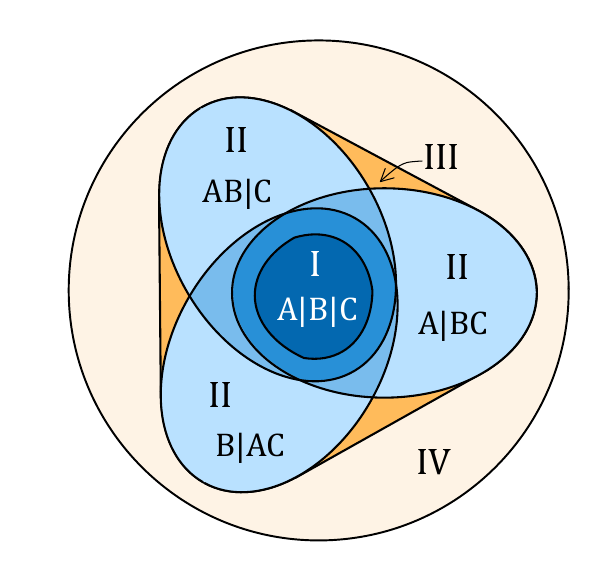}
    \caption{Graphical representation of sets of tripartite states with different separability properties. The convex hull of partition-separable states (II) forms the set of biseparable states, and its complement to the whole set of states is composed of states with GME (IV). The complement to the set of partition-separable states within the set of biseparable states is the set of FIB states (III). The union of the sets (III) and (IV) forms the set of fully inseparable states.}
    \label{fig:entanglementClasses}
\end{figure}

{\noindent}We consider a quantum state shared by three parties labelled $A$, $B$, and $C$, and the labels can be collected in a set $\mathcal{M}=\{A, B, C\}$. The set $\mathcal{M}$ can be split into three different pairs of disjoint nonempty subsets $I_k=\{k\}$, $k=A, B, C$, and its complement $\overline{I}_k=\mathcal{M}\setminus I_k$ for $k=A, B, C$, where the different splittings are referred to as \textit{bipartitions} $k|lm$, $k,l,m\in\{A,B,C\}$ with $k\neq l\neq m\neq k$. The state is separable with respect to the bipartition $k|lm$, if it can be expressed as~\cite{Werner1989, HorodeckiEntanglementReview2009}
\begin{equation}\label{eq:bipartite}
    \rho^{\mathrm{sep}}_{k|lm}=\sum_jp_j\rho_{k}^{(j)}\otimes\rho_{lm}^{(j)},
\end{equation}
with the $p_j$ satisfying $0\leq p_j \leq 1$ and $\sum_jp_j=1$ being probability weights, and we refer to these states as \textit{partition-separable states}. States that are not separable with respect to any bipartition are referred to as \textit{fully inseparable states}. By convex mixing of states that can be written as in Eq.~(\ref{eq:bipartite}), we obtain a set of all \emph{biseparable} states~\cite{BancalGisinLiangPironio2011}, i.e., states that can be written as
\begin{equation}\label{eq:biseparable}
    \rho^{\mathrm{bisep}}=p_1\rho^{\mathrm{sep}}_{A|BC}+p_2\rho^{\rm sep}_{B|AC}+p_3\rho^{\rm sep}_{AB|C}
\end{equation}
with $p_1+p_2+p_3=1$ and $p_1,p_2,p_3\geq0$. Full inseparability and GME are inequivalent notions due to the existence of fully inseparable biseparable (FIB) states that can be decomposed into the convex combination of partition-separable states, while states that are fully inseparable but cannot be written in the form~(\ref{eq:biseparable}) are called genuinely multipartite entangled~\cite{GuehneToth2009, FriisVitaglianoMalikHuber2019, BertlmannFriis2023}. The structure of the set of quantum states of three parties is illustrated in Fig.~\ref{fig:entanglementClasses}.

\subsection{Entanglement detection}\label{subsec:detectionEnt}

{\noindent}In this subsection, we briefly explain the entanglement and GME detection methods used in this paper.

To detect entanglement across a fixed bipartition of the state, the positive partial-transposition (PPT) criterion~\cite{HorodeckiMPR1996,Peres1996} can be used: For all states $\rho$ that are separable with respect to a chosen partition $k|lm$, the partially transposed density operator $(\top_{k}\otimes\mathds{1}_{lm})[\rho]$ is positive semi-definite. Here $\top_{k}$ stands for transposition over mode $k$. However, states that satisfy this criterion, so-called PPT states, need not be separable. For CV systems the PPT criterion can be cast in the form of a criterion for the second moments. Specifically, the CM of any state that is separable across a bipartition $k|lm$ must obey the relation~\cite{Simon2000}
\begin{equation}\label{eq:PPT}
    \gamma^{\top\subtiny{0}{0}{k}}+i\Omega\geq0,
\end{equation}
where the matrix~$\Omega$ is given in Eq.~(\ref{eq:omega3}). The CM of the partially transposed state $\rho\suptiny{0}{0}{\top_k}$ is given by the relation $\gamma\suptiny{0}{0}{\top_k}=(\sigma_{Z,k}\oplus\mathds{1}_{lm})\gamma (\sigma_{Z,k}\oplus\mathds{1}_{lm})$, where $\sigma_{Z,k}=\mathrm{diag}(1,-1)$ is Pauli $Z$-matrix on mode $k$. The violation of~(\ref{eq:PPT}) gives a sufficient condition for entanglement with respect to the bipartition $k|lm$, while for Gaussian states, the condition is also necessary~\cite{WernerWolf2001}.

To detect GME, it is common (see, e.g.,~\cite{GuehneToth2009,FriisVitaglianoMalikHuber2019}) to use a witness operator $W$, which is a Hermitian operator that satisfies $\Tr(W\rho^{\rm bisep})\geq0$ for all biseparable states and $\Tr(W\rho)<0$ for at least one genuinely multipartite entangled state. For discrete-variable systems, this can be realized, for instance, by a witness known as a fully decomposable witness (or PPT mixer in other literature)~\cite{JungnitschMoroderGuehne2011a}. This witness distinguishes convex mixtures of PPT states for different bipartitions~\cite{HorodeckiMPR1996}, which includes all biseparable states, from (a subset of) genuinely multipartite entangled states with non-positive partial transposition (NPT states) for all bipartitions. However, we note that there are some genuinely multipartite entangled states (that are either PPT for one or more bipartitions, or NPT for all bipartitions but can be written as convex combinations of PPT states for different bipartitions) that are not detected by this witness. The search for the fully decomposable witness can be formulated as a semidefinite program (SDP)~\cite{JungnitschMoroderGuehne2011a} given by
\begin{align}
    & \underset{W,\{P_{k}, Q_k\}}{\text{minimize}}
    & & \Tr(W\rho\subtiny{0}{0}{ABC}) \nonumber\\
    & \text{subject to}
    & & \Tr (W) = 1,\label{app-witness:dual}\\
    & & & W = P_k+Q_k^{\top_k}\geq 0, \quad \textrm{for} \quad  k = \{A, B, C \},\nonumber\\
    & & & P_{k}\geq0,\quad Q_{k}\geq0.\nonumber
\end{align}
Another possibility for the detection of GME from the density matrix $\rho$ of the state is based on comparisons of off-diagonal and diagonal elements of the density matrix. The criterion from Ref.~\cite{GabrielHiesmayrHuber2010} gives a necessary condition for biseparability of the state $\rho$, and in the general form it can be written as
\begin{align}\label{app-criterion:Gabriel}
    &\sqrt{\brakket{\phi}{\rho^{\otimes2}P_{\rm tot}}{\phi}}\nonumber\\
    &\quad\leq \sum_{k=A,B,C}\left(\brakket{\phi}{P^\dagger_{k}\rho^{\otimes2}P_{k}}{\phi}\brakket{\phi}{P^\dagger_{lm}\rho^{\otimes2}P_{lm}}{\phi}\right)^{1/4},
\end{align}
where $\ket{\phi}$ is a fully separable tripartite state on the Hilbert space of two copies of $\rho$, $P_{\rm tot}$ is an operator exchanging the two copies, and $P_{k}(P_{lm})$ is an operator exchanging the parts of the copies associated with the subsystem $k$ (subsystems $lm$). A generalized version of the criterion~(\ref{app-criterion:Gabriel}) was introduced in~\cite[Eq.~(A.4)]{FriisHuberFuentesBruschi2012} and further extended in~\cite{BaksovaLeskovjanovaMistaAgudeloFriis2025} to formulate a necessary condition of biseparability as
\begin{align}\label{app-criterion:advancedGabriel}
    &\abs{\brakket{000}{\rho}{011}}+\abs{\brakket{000}{\rho}{101}}+\abs{\brakket{000}{\rho}{110}}\nonumber\\[1mm]
    & \leq\sqrt{\brakket{000}{\rho}{000}}\sqrt{\brakket{011}{\rho}{011}+\brakket{101}{\rho}{101}+\brakket{110}{\rho}{110}}\nonumber\\[1mm]
    &\quad +\sqrt{\brakket{001}{\rho}{001}\brakket{010}{\rho}{010}}+\sqrt{\brakket{001}{\rho}{001}\brakket{100}{\rho}{100}}\nonumber\\[1mm]
    &\quad +\sqrt{\brakket{010}{\rho}{010}\brakket{100}{\rho}{100}}.
\end{align}
The violation of the inequalities (\ref{app-criterion:Gabriel}) or (\ref{app-criterion:advancedGabriel}) implies GME in the investigated state. While a plethora of other detection methods for GME exists, many of these are tailored to the detection of specific finite-dimensional target states (including graph or stabilizer states~\cite{TothGuehne2005b, AudenaertPlenio2005, SmithLeung2006, HeinDuerEisertRaussendorfVanDenNestBriegel2006, JungnitschMoroderGuehne2011b, ZhouZhaoYuanMa2019, LiDaiMunozAriasReuerHuberFriis2026}, GHZ and W like states~\cite{GuehneSeevinck2010, HuberMintertGabrielHiesmayr2010}, Dicke states~\cite{BergmannGuehne2013}, or states with specific symmetries~\cite{TothGuehne2009}), are based on specific measurement restrictions (e.g., access to only two-body~\cite{FriisMartyEtal2018, ParaschivMiklinMoroderGuehne2018} or certain few-body correlators~\cite{CanteriBateMishraFriisKrutyanskiyLanyon2025,LiDaiMunozAriasReuerHuberFriis2026}), or built to extract other information (such as, on high-dimensional entanglement in terms of the Schmidt-number vector~\cite{HuberDeVicente2013, HuberPerarnauDeVicente2013, LiuHeHuberVitagliano2026}), and are therefore not of immediate use to us here.

In  CV systems, one can make use of a criterion introduced in Ref.~\cite{HyllusEisert2006} that we call the CM biseparability criterion, which states that the CM $\gamma^{\rm bisep}$ of any biseparable state must satisfy
\begin{equation}\label{eq:CMbisepDecomp}
    \gamma^{\rm bisep}-(p_1\gamma_A\oplus\gamma_{BC}+p_2\gamma_{AC}\oplus\gamma_B+p_3\gamma_{AB}\oplus\gamma_C)\geq0,
\end{equation}
for some CMs $\gamma_A$, $\gamma_B$, $\gamma_C$, $\gamma_{BC}$, $\gamma_{AC}$, and $\gamma_{AB}$, and probabilities $p_1,p_2,p_3\geq0$ with $p_1+p_2+p_3=1$. The violation of~(\ref{eq:CMbisepDecomp}) implies GME in the state with the considered CM. To test if a given CM satisfies the relation~(\ref{eq:CMbisepDecomp}), we can use a witness matrix $Z$, which is a real symmetric positive semi-definite matrix for which $\Tr(Z\gamma^{\rm bisep})\geq1$ for all CMs satisfying~(\ref{eq:CMbisepDecomp}) and $\Tr(Z\gamma)<1$ for at least one CM vioaliting~(\ref{eq:CMbisepDecomp}). To find an optimal matrix $Z$, one can use an SDP introduced in~\cite{HyllusEisert2006} and publicly available in~\cite{ProvaznikGithub}.

\section{GME in Gaussian states}\label{sec:results}

We now focus on the detection of GME in three-mode Gaussian states whose CM $\gamma$ satisfies inequality~(\ref{eq:CMbisepDecomp}). 

While the CM of every biseparable state (\ref{eq:biseparable}) satisfies the condition (\ref{eq:CMbisepDecomp}), a Gaussian state with CM satisfying this condition can be genuinely multipartite entangled, as was shown in~\cite{BaksovaLeskovjanovaMistaAgudeloFriis2025}.
This means that the GME in such Gaussian states cannot be detected by standard GME criteria that are based solely on second moments of arbitrary states~\cite{HyllusEisert2006, TehReid2014, ShchukinLoock2015, ToscanoSaboiaAvelarWalborn2015}. This is because for every CM satisfying~(\ref{eq:CMbisepDecomp}), one can always find a non-Gaussian biseparable state. Therefore, any such criterion on second moments that would detect GME in a Gaussian state with CM satisfying the condition (\ref{eq:CMbisepDecomp}) would also detect GME in biseparable states.

An alternative strategy is to use criteria based on higher moments such as those in Refs.~\cite{ShchukinLoock2014, ZhangBarralZhangXiaoBencheikh2023}, as higher moments differ for Gaussian and non-Gaussian states with the same CM. However, these criteria do not detect GME in the example from~\cite{BaksovaLeskovjanovaMistaAgudeloFriis2025}.

Therefore, we employ a similar strategy as in~\cite{BaksovaLeskovjanovaMistaAgudeloFriis2025}, and focus on local projections of the density matrix of the Gaussian state whose CM satisfies the condition~(\ref{eq:CMbisepDecomp}) to finite-dimensional subspaces, where we can use GME criteria on projected discrete-variable states (qudits). As the projection to qudits is a local operation, and therefore cannot create entanglement, we know that the original Gaussian state is genuinely multipartite entangled at least in the same region where the respective projected qudit state is GME. Below, we present results for different families of Gaussian states that are fully inseparable (according to the CV version of the PPT criterion~\cite{Simon2000,WernerWolf2001}) and whose CM satisfies the biseparability criterion of~(\ref{eq:CMbisepDecomp}). 
In the first part of this analysis, we take families of FIB non-Gaussian states and study Gaussian states associated with them by virtue of giving rise to the same CM, which fulfills the condition (\ref{eq:CMbisepDecomp}) by construction.
In the second part, we investigate Gaussian states defined by a CM obtained from mixing the CM of a pure fully inseparable Gaussian state with that of a pure fully separable Gaussian state, and find regions of its parameters where it satisfies decomposition~(\ref{eq:CMbisepDecomp}) using the SDP from Refs.~\cite{HyllusEisert2006, ProvaznikGithub}.

\subsection{Two-mode squeezed vacuum state with single-mode state}

{\noindent}We start with four families of FIB non-Gaussian states that are defined as convex mixtures of three-mode states that are each partition separable with respect to one partition and entangled with respect to the other two partitions. Specifically, these three-mode states are product states comprised of a TMSV state for two of the modes and some single-mode state for the remaining mode, such that the total states are
\begin{equation}\label{rho:symmetric}
    \rho\suptiny{0}{0}{(i)}\subtiny{0}{0}{ABC}=\tfrac{1}{3}\left(\rho\suptiny{0}{0}{\rm TMSV}\subtiny{0}{0}{AB}\otimes\rho\suptiny{0}{0}{(i)}\subtiny{0}{0}{C}+\rho\suptiny{0}{0}{\rm TMSV}\subtiny{0}{0}{AC}\otimes\rho\suptiny{0}{0}{(i)}\subtiny{0}{0}{B}+\rho\suptiny{0}{0}{(i)}\subtiny{0}{0}{A}\otimes\rho\suptiny{0}{0}{\rm TMSV}\subtiny{0}{0}{BC}\right).
\end{equation}
Here, we use the label ``$i$" to distinguish between three-mode states with different choices of single-mode states $\rho\suptiny{0}{0}{(i)}$, while the density matrix of the TMSV is given by
\begin{equation}\label{app-rho:TMSV}
    \rho\suptiny{0}{0}{\mathrm{TMSV}}\subtiny{0}{0}{jk}=(1-\lambda^2)\sum_{n,n'=0}^\infty\lambda^{n+n'}\ket{nn}\!\!\bra{n'n'}\subtiny{0}{0}{jk},
\end{equation}
where $\lambda=\tanh{(r)}$ with $r\geq0$ is the squeezing parameter, and $\ket{n}\subtiny{0}{0}{j}$ is a Fock state of mode $j$ obtained by $n$-fold application of the creation operator $a\subtiny{0}{0}{j}^{\dagger}$ to the vacuum $\ket{0}$, that is, $\ket{n}=(1/\sqrt{n!})(a\subtiny{0}{0}{j}^{\dagger})^{n}\ket{0}$. 

When the state (\ref{rho:symmetric}) has zero first moments, its CM is in the form
\begin{equation}\label{gamma:symmetric}
    \gamma\subtiny{0}{0}{ABC}\suptiny{0}{0}{(i)}=\tfrac{1}{3}\left(\gamma\subtiny{0}{0}{AB}\suptiny{0}{0}{\mathrm{TMSV}}\!\oplus\!\gamma\subtiny{0}{0}{C}\suptiny{0}{0}{(i)}
    \!+\gamma\subtiny{0}{0}{AC}\suptiny{0}{0}{\mathrm{TMSV}}\!\oplus\!\gamma\subtiny{0}{0}{B}\suptiny{0}{0}{(i)}
    \!+\gamma\subtiny{0}{0}{A}\suptiny{0}{0}{(i)}\oplus\gamma\subtiny{0}{0}{BC}\suptiny{0}{0}{\mathrm{TMSV}}\right),
\end{equation}
where the CM of the TMSV state is
\begin{equation}\label{app-cm:TMSV}
    \gamma\suptiny{0}{0}{\mathrm{TMSV}}\subtiny{0}{0}{jk}=\left(\begin{array}{cc}
        a  & c \\[2mm]
        c & a
    \end{array}
    \right)
    \oplus\left(\begin{array}{cc}
        a  & -c \\[2mm]
        -c & a
    \end{array}
    \right),
\end{equation}
with $a=\cosh(2r)$ and $c=\sinh(2r)$.

With this definition at hand, we first follow up on the example from~\cite{BaksovaLeskovjanovaMistaAgudeloFriis2025}, where the single-mode state (for $i=0$) is a vacuum state, $\rho\suptiny{0}{0}{(0)}=\ket{0}\!\bra{0}$. As the vacuum state has zero first moments, the resulting CM has the form given by Eq.~(\ref{gamma:symmetric}) with $\gamma\subtiny{0}{0}{j}\suptiny{0}{0}{(0)}=\mathbb{1}$.
In~\cite{BaksovaLeskovjanovaMistaAgudeloFriis2025}, the CM $\gamma\subtiny{0}{0}{ABC}\suptiny{0}{0}{(0)}$ was detected as fully inseparable for the range $0<r<r_0$ of the squeezing parameter with $r_0\approx1.24275$, and the corresponding Gaussian state $\rho\subtiny{0}{0}{\mathrm{G}}\suptiny{0}{0}{(0)}$ was detected by the criterion~(\ref{app-criterion:advancedGabriel}) as genuinely multipartite entangled for the parameter range $0<r<r_1$ with $r_1=0.284839$, and by the fully decomposable witness applied to the local projection of the state $\rho\subtiny{0}{0}{\mathrm{G}}\suptiny{0}{0}{(0)}$ to three qubits for the parameter range $0<r<r_2$ with $r_2\approx0.575584$.

\begin{figure}[h]\label{fig:TMSVvac}
    \centering
    \includegraphics[width=1\linewidth]{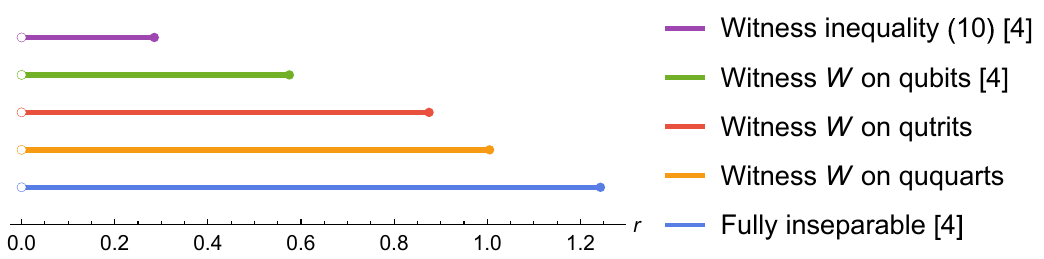}
    \caption{Comparison of ranges of the threshold squeezing parameter~$r$ for which the Gaussian state with CM $\gamma\subtiny{0}{0}{ABC}\suptiny{0}{0}{(0)}$ is detected as fully inseparable (blue, bottom), and as genuinely multipartite entangled using different detection methods (top to bottom): the witness inequality (\ref{app-criterion:advancedGabriel}), or the fully decomposable witness in the form of the SDP in (\ref{app-witness:dual}) after local projection to three qubits (green, second from the top)~\cite{BaksovaLeskovjanovaMistaAgudeloFriis2025}, to three qutrits (red, third from the top), and to three ququarts (yellow, fourth from the top).}
\end{figure}

We extend this investigation by locally projecting the state $\rho\subtiny{0}{0}{\mathrm{G}}\suptiny{0}{0}{(0)}$ to subspaces of larger local dimensions, namely to three-qutrit ($d=3$) and three-ququart ($d=4$) subspaces. To do this, we use methods explained in more detail in Appendix~\ref{app-sec:DMmapping} to obtain matrix elements with respect to the Fock states $\ket{0}$, $\ket{1}$, $\ket{2}$, and $\ket{3}$, and test for GME using the fully decomposable witness described in Sec.~\ref{subsec:detectionEnt}. Due to the computational cost of the SDP, we vary the parameter~$r$ in increments of $0.005$.
 The qutrit state was detected as genuinely multipartite entangled for the parameter range $0<r\leq0.875$ and the ququart state for the range $0<r\leq1.005$. One can see that the range of~$r$ where the state is detected as genuinely multipartite entangled increases and approaches the border of full inseparability with increasing dimension of the projected state (see Fig.~\ref{fig:TMSVvac}).

We continue by selecting density matrices $\rho\suptiny{0}{0}{(i)}$ for $i=1,2,3$ to construct the conrresponding non-Gaussian states described by Eq.~(\ref{rho:symmetric}) and the associated CMs~$\gamma\suptiny{0}{0}{(i)}$. Specifically, $\rho\suptiny{0}{0}{(i)}$ is selected as a single-mode squeezed vacuum state, a thermal state, and a coherent state, for $i=1$, $i=2$, and $i=3$, respectively. We then investigate GME in the corresponding Gaussian states with the same CMs~$\gamma\suptiny{0}{0}{(i)}$.

\subsubsection*{(1) Single-mode squeezed vacuum state}

{\noindent}Firstly, we replace the vacuum state from the previous example with a more general single-mode squeezed vacuum state (SMSV) where we use the same symbol ($r$) for the squeezing parameter as for the TMSV to obtain a one-parameter family of states. The resulting CM is given by Eq.~(\ref{gamma:symmetric}) with the CM of the SMSV state given by
\begin{align}
    \gamma\subtiny{0}{0}{j}\suptiny{0}{0}{(1)}=\gamma\suptiny{0}{0}{\mathrm{SMSV}}\subtiny{0}{0}{j}=\left(\begin{array}{cc}
       e^{2r}  & 0\\
        0 & e^{-2r}
    \end{array}
    \right).\label{app-cm:SMSV}
\end{align}

\begin{figure}[h]
    \centering
    \includegraphics[width=1\linewidth]{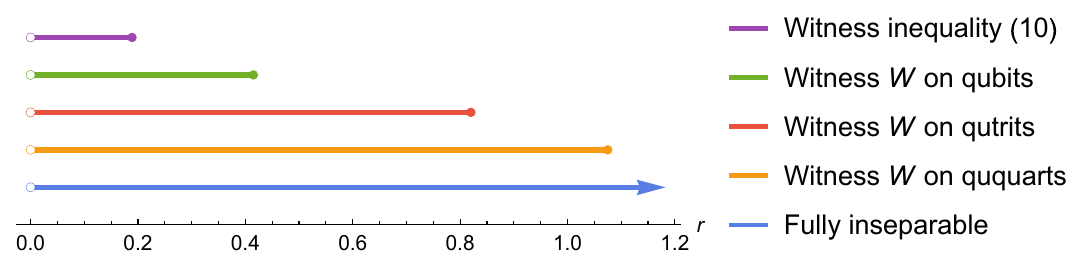}
    \caption{Comparison of ranges of the squeezing parameter~$r$ for which the Gaussian state with CM $\gamma\subtiny{0}{0}{ABC}\suptiny{0}{0}{(1)}$ is detected as fully inseparable (blue, bottom), and as genuinely multipartite entangled using different detection methods (top to bottom): the witness inequality (\ref{app-criterion:advancedGabriel}), or the fully decomposable witness in the form of the SDP in (\ref{app-witness:dual}) after local projection to three qubits (green, second from the top), to three qutrits (red, third from the top), and to three ququarts (yellow, fourth from the top).}
    \label{fig:TMSVSMSV}
\end{figure}

Since the state~(\ref{rho:symmetric}) is an equally weighted mixture of partition-separable states that are the same up to a permutation of the modes, the state is symmetric under exchange of every pair of modes, and to find its fully inseparable region, we thus only need to examine one bipartition. 
We hence numerically find that the CM $\gamma\subtiny{0}{0}{ABC}\suptiny{0}{0}{(1)}$ is fully inseparable for $r>0$. We calculated the density-matrix elements of the corresponding Gaussian state $\rho\subtiny{0}{0}{\mathrm{G}}\suptiny{0}{0}{(1)}$ in the three-mode Fock basis $\{\ket{m,n,k}\}_{m,n,k}$ for $m,n,k=0,1,2,3$. The criterion~(\ref{app-criterion:Gabriel}) does not detect GME in these states, while the criterion~(\ref{app-criterion:advancedGabriel}) detects GME for $0<r<0.1893$, and the fully decomposable witness based on the SDP in~(\ref{app-witness:dual}) detects GME after local projection to three qubits (local dimension $d=2$) for $0<r\leq 0.4172$, to qutrits ($d=3$) for $0<r\leq0.820$, and for ququarts ($d=4$) for $0<r\leq1.075$. One can see that, similarly to the case of $\gamma\subtiny{0}{0}{ABC}\suptiny{0}{0}{(0)}$, the detected region increases with increasing local dimension~$d$ of the subspaces on which the state is projected, while the detected region grows noticably faster with increasing~$d$ than in the previous case (see Fig.~\ref{fig:TMSVSMSV}).

\subsubsection*{(2) Thermal state}

{\noindent}As the next example, we consider a mixed state that is combined with the TMSV state in Eq.~(\ref{rho:symmetric}), which is given by a thermal state
\begin{equation}
    \rho\suptiny{0}{0}{(2)}\subtiny{0}{0}{j}=\frac{1}{1+\overline{n}}\sum_{m=0}^\infty\left(\frac{\overline{n}}{1+\overline{n}}\right)^m\ketbra{m}{m},
\end{equation}
where~$\overline{n}$ is the mean photon number of the thermal state. The resulting CM of the mixture is given by Eq.~(\ref{gamma:symmetric})
where
\begin{align}
    \gamma\subtiny{0}{0}{j}\suptiny{0}{0}{(2)}=\gamma\suptiny{0}{0}{\mathrm{th}}\subtiny{0}{0}{j}=\left(\begin{array}{cc}
       2\overline{n}+1  & 0\\
        0 & 2\overline{n}+1
    \end{array}
    \right).\label{cm:thermal}
\end{align}

\begin{figure}[t]
    \centering
    \includegraphics[width=0.95\linewidth]{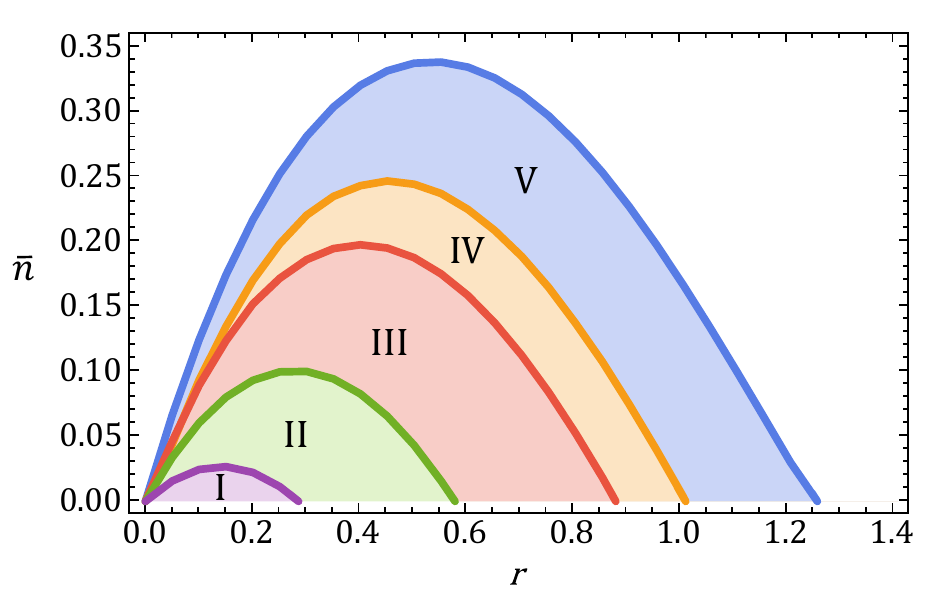}
    \caption{Comparison of regions where the Gaussian state with CM $\gamma\suptiny{0}{0}{(2)}\subtiny{0}{0}{ABC}$ is detected as GME by the witness inequality~(\ref{app-criterion:advancedGabriel}) (I), and by the fully decomposable witness \textit{W} on qubits (II), qutrits (III), and ququarts (IV), inside the fully-inseparable region (V).}
    \label{app-fig:TMSVthermal}
\end{figure}

We apply the same methods as in the previous case, and the resulting regions of full inseparability, GME detection via the witness inequality~(\ref{app-criterion:advancedGabriel}), and GME detection via the fully decomposable witness $W$ are shown in Fig.~\ref{app-fig:TMSVthermal}.
There, the squeezing parameter~$r$ varies in increments of 0.05, and for each value of the parameter $r$, we iteratively searched for the maximum detected value of the parameter $\overline{n}$ up to a precision of $0.0001$ for qubits and qutrits, and $0.0003$ for ququarts.
Linear interpolations of these values create boundaries of the green (II) region for qubits ($d = 2$), red region (III) for qutrits ($d = 3$), and orange (IV) region for ququarts ($d = 4$).

\subsubsection*{(3) Coherent state}

{\noindent}As the last example, we also analyse a state with non-zero first moments. Specifically, we combine the TMSV states with a coherent state $\rho\suptiny{0}{0}{(i)}\subtiny{0}{0}{j}=\ket{\alpha}\!\!\bra{\alpha}\subtiny{0}{0}{j}$. The CM of the Gaussian state cannot be directly written as in Eq.~(\ref{gamma:symmetric}) due to non-zero first moments of the coherent state. The resulting CM is of the form
\begin{equation}\label{app-cm:TMSVcoherent}
    \gamma\subtiny{0}{0}{ABC}\suptiny{0}{0}{(3)}=\frac{1}{9}\left(\begin{array}{ccc}
        a & b & b \\
        b & a & b \\
        b & b & a
    \end{array}\right)\oplus\frac{1}{3}\left(\begin{array}{ccc}
        c & d & d \\
        d & c & d \\
        d & d & c
    \end{array}\right),
\end{equation}
where $a=8\alpha^2+6\cosh(2r)+3$, $b=3\sinh(2r)-4 \alpha^2$, $c=2\cosh(2r)+1$, and $d=-\sinh(2r)$. The parameter~$\alpha$ of the coherent state is chosen to be real and nonnegative (as the state is symmetric around $\alpha=0$).

\begin{figure}[t]
    \centering
    \includegraphics[width=0.9\linewidth]{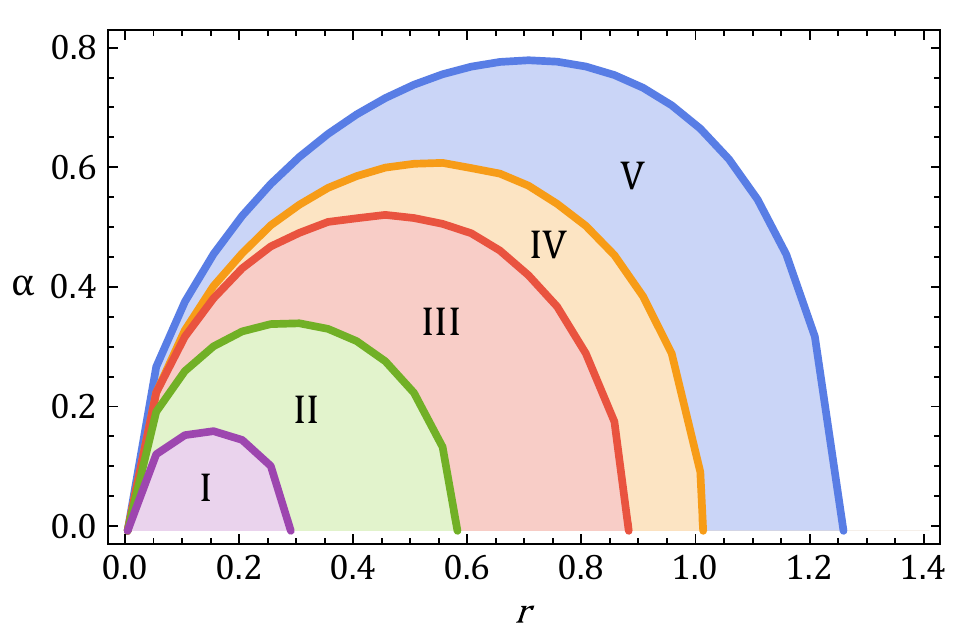}
    \caption{Comparison of regions where the Gaussian state with CM $\gamma\suptiny{0}{0}{(3)}\subtiny{0}{0}{ABC}$, Eq.~(\ref{app-cm:TMSVcoherent}), is detected as GME by the witness inequality~(\ref{app-criterion:advancedGabriel}) (I), and by the fully decomposable witness \textit{W} on qubits (II), qutrits (III), and ququarts (IV), inside the fully-inseparable region (V).}
    \label{app-fig:TMSVcoherent}
\end{figure}

The CM~(\ref{app-cm:TMSVcoherent}) violates the CV PPT criterion (\ref{eq:PPT}) for parameters
\begin{equation}\label{app-border:TMSVcoherentNPT-part1}
    0\ \textcolor{darkgreen}{<}\ r< 1.2428
\end{equation}
and
\begin{equation}\label{app-border:TMSVcoherentNPT-part2}
    0\leq \alpha <\frac{1}{2}\sqrt{\frac{66\sinh(r)+31\sinh(3r)-3\sinh(5r)}{22\cosh(r)+2\cosh(3r)+4\sinh(3r)}},
\end{equation}
and therefore it is fully inseparable for this range of parameters (see blue region V in Fig.~\ref{app-fig:TMSVcoherent}).

We calculated the density-matrix elements of the corresponding Gaussian state $\rho\subtiny{0}{0}{\mathrm{G}}\suptiny{0}{0}{(3)}$ in the three-mode Fock basis $\{\ket{m,n,k}\}_{m,n,k}$ for $m,n,k=0,1,2,3$ and tested GME using the criterion~(\ref{app-criterion:advancedGabriel}) and the fully decomposable witness obtained from the SDP~(\ref{app-witness:dual}) on qubits, qutrits, and ququarts.
The results are illustrated in Fig.~\ref{app-fig:TMSVcoherent}, where the regions are given by linear interpolations of bordering pairs of the parameter~$\alpha$ and the squeezing parameter~$r$. As in the previous case, the squeezing parameter~$r$ varies in increments of 0.05,
and for each value of the parameter $r$, we iteratively searched for the maximum detected value of the parameter $\alpha$ up to a precision of $0.0001$ for qubits and qutrits, and $0.0008$ for ququarts.

One can notice that the regions for which GME is detected depend on the parameter~$\alpha$. However, this is not in contradiction with the fact that entanglement is not affected by displacement. It arises because we mix states with different first moments, which adds classical noise to the covariance matrix.

\subsection{Noisy GHZ-like state}\label{sec:GHZ}

{\noindent}It is also appropriate to test GME for a Gaussian state that is not derived from the biseparable convex mixture~(\ref{rho:symmetric}). For this purpose, we choose a three-mode GHZ-like state~\cite{HyllusEisert2006} with CM $\gamma\suptiny{0}{0}{\mathrm{GHZ}}$, which has the form
\begin{equation}\label{cm:GHZ}
    \gamma\suptiny{0}{0}{\mathrm{GHZ}}=\left(\begin{array}{ccc}
        a_+ & c_+ & c_+ \\
        c_+ & a_+ & c_+ \\
        c_+ & c_+ & a_+
    \end{array}\right)\oplus\left(\begin{array}{ccc}
        a_- & c_- & c_- \\
        c_- & a_- & c_- \\
        c_- & c_- & a_-
    \end{array}\right),
\end{equation}
where
\begin{equation}\label{eq:GHZparameters}
    a_\pm=\frac{1}{3}\left(e^{\pm 2r}+2e^{\mp 2r}\right)\;,\quad c_\pm=\frac{1}{3}\left(e^{\pm 2r}-e^{\mp 2r}\right),
\end{equation}
and $r\geq 0$ is the squeezing parameter. The GHZ-like state with CM $\gamma\suptiny{0}{0}{\mathrm{GHZ}}$ is fully inseparable according to criterion~(\ref{eq:PPT}) for $r>0$, and since it is a pure state, it is also genuinely multipartite entangled in this range of squeezing parameters, as the set of fully inseparable states and the set of genuinely multipartite entangled states coincide for pure states.

\begin{figure}[t]
    \centering
    \includegraphics[width=1\linewidth]{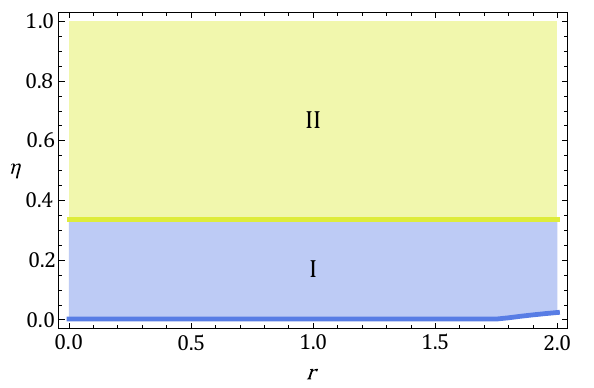}
    \caption{Regions of partition separability (small white region in the lower right corner), full inseparability (I), and GME (II) of the state given by CM witnesses in the state with CM Eq.~(\ref{cm:noisyGHZ}).}
    \label{fig:noisyGHZ-CMresults}
\end{figure}

In the following, we investigate GME in a noisy GHZ-like state, which we define as a mixed Gaussian state with CM $\gamma\suptiny{0}{0}{(4)}$ obtained by mixing the CM of the GHZ-like state with the CM of a vacuum state, i.e.,
\begin{equation}\label{cm:noisyGHZ}
    \gamma\suptiny{0}{0}{\mathrm{(4)}}=\eta\gamma\suptiny{0}{0}{\mathrm{GHZ}}+(1-\eta)\mathbb{1},
\end{equation}
where $0\leq\eta\leq1$ and $\mathbb{1}$ is the $6\times6$ identity matrix. The regions of parameters $(r,\eta)$ where the state is partition separable and thus fulfills the condition (\ref{eq:PPT}) are given by 
\begin{equation}\label{border:fs}
    r>\frac{1}{2}\log\left(17+12\sqrt{2}\right)\; \wedge \; 0\leq\eta\leq 1-\frac{2\sqrt{2}}{3}\coth{(r)},
\end{equation}
and the regions where the state reduces to the vacuum state, i.e.~$r=0$ or~$\eta=0$. The region given by Eq.~(\ref{border:fs}) can be seen in Fig.~\ref{fig:noisyGHZ-CMresults} as the small white region in the bottom right corner.

To find parameters for which the Gaussian state with CM~(\ref{cm:noisyGHZ}) satisfies the condition~(\ref{eq:CMbisepDecomp}) we used the SDP from~\cite{HyllusEisert2006,ProvaznikGithub}. We find that the state does not obey the condition~(\ref{eq:CMbisepDecomp}) and therefore is genuinely multipartite entangled for $\eta>\frac{1}{3}$ and $r\in\left(0,2\right]$, where the upper bound on the squeezing parameter was chosen to be higher than standard squeezing that can be realized in current state-of-the-art experiments \cite{VahlbruchMehmetDanzmannSchnabel2016} (see Fig.~\ref{fig:noisyGHZ-CMresults}).

\begin{figure}[t]
    \centering
    \includegraphics[width=1\linewidth]{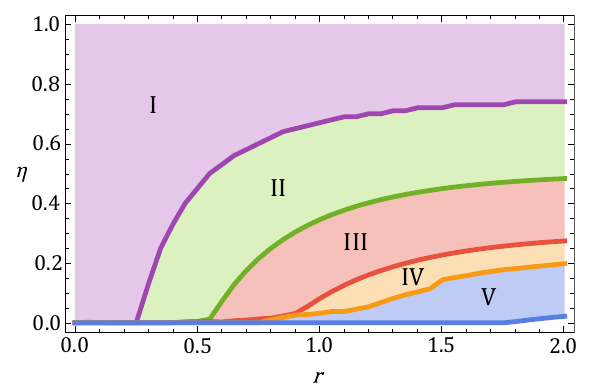}
    \caption{Comparison of regions where the Gaussian state with CM $\gamma\suptiny{0}{0}{(4)}\subtiny{0}{0}{ABC}$, Eq.~(\ref{cm:noisyGHZ}), is detected as GME by the witness inequality~(\ref{app-criterion:advancedGabriel}) (I), and by the fully decomposable witness \textit{W} on qubits (II), qutrits (III), and ququarts (IV), inside the fully-inseparable region (V).}
    \label{fig:noisyGHZcomparison}
\end{figure}

We now wish to identify parameter regions for which the corresponding states are fully inseparable and condition~(\ref{eq:CMbisepDecomp}) is satisfied, but where other criteria detect GME. To this end, we again calculate the density-matrix elements in the three-mode Fock basis $\{\ket{m,n,k}\}_{m,n,k}$ for $m,n,k=0,1,2,3$, and apply criteria~(\ref{app-criterion:Gabriel}) and~(\ref{app-criterion:advancedGabriel}) as well as the fully decomposable witness to the corresponding qubit, qutrit, and ququart states. As in the previous cases, the criterion from~(\ref{app-criterion:Gabriel}) does not detect GME in these states. In contrast, the criterion from~(\ref{app-criterion:advancedGabriel}) successfully detects GME in some states, shown as the purple (I) region in Fig.~\ref{fig:noisyGHZcomparison}. For the fully decomposable witness, the squeezing parameter~$r$ is varied in increments of $0.05$. For each squeezing~$r$, we determine the minimal value of the noise parameter~$\eta$ for which the state is detected as genuinely multipartite entangled for a given local dimension. The parameter~$\eta$ varies with fixed increments over several iterations, where in each iteration the increment 
is decreased with respect to 
the previous iteration. In the last iteration, the increment was $0.0001$ for qubits and qutrits, and $0.001$ for ququarts. Linear interpolations of these values create boundaries of the green (II) region for qubits ($d=2$), red region (III) for qutrits ($d=3$), and orange (IV) region for ququarts ($d=4$) shown in Fig.~\ref{fig:noisyGHZcomparison}.

We observe that all detection methods based on local projections to finite-dimensional subspaces outperform the CM condition~(\ref{eq:CMbisepDecomp})
(yellow region II in Fig.~\ref{fig:noisyGHZ-CMresults}) for small squeezing. For larger squeezing, however, the CM condition becomes more effective than the criterion from~(\ref{app-criterion:advancedGabriel}) and the fully decomposable witness applied to qubits.
Overall, the results provide further indication that the gap between genuinely multipartite entangled states and partition-separable states decreases as the considered truncation dimension increases.

Across all examined families, we observe that Gaussian states with CMs permitting a decomposition of the form of~(\ref{eq:CMbisepDecomp}) are detected as genuinely multipartite entangled for some range of parameters. While these results do not rule out the existence of parameter regions where the states are FIB, the region where the state is fully inseparable but not detected as genuinely multipartite entangled shrinks with increasing local dimension of the subspace to which the Gaussian state is projected.

Furthermore, we are not aware of a way to construct an example of a FIB Gaussian state. Indeed, one can already observe that mixtures of distinct univariate Gaussian distributions are not Gaussian~\cite{Teicher1960, Teicher1963}.

Altogether, this suggests that FIB Gaussian states may not exist. Therefore, we formulate the following conjecture:\\

\noindent\textbf{Conjecture:} \textit{All fully inseparable Gaussian states are genuinely multipartite entangled.}\\

\section{Discussion and conclusion}\label{sec:Conclussion}

{\noindent}In this paper, we have investigated GME in fully inseparable Gaussian states whose CM corresponds to (non-Gaussian) states with biseparable decomposition, which rules out the detection of GME in such states by criteria based on the CM of an arbitrary state. We presented several families of Gaussian states whose CMs fulfill these conditions, calculated their density-matrix elements with respect to the Fock basis for up to three excitations (local dimension $d=4$), and then employed GME criteria for density matrices of finite-dimensional systems. We observe that the detection of GME in Gaussian states with these CMs is common, and for several examples of such states, we show that the parameter regions of fully inseparable states that are not detected as genuinely multipartite entangled shrink with increasing dimension of the subspaces that the states are projected onto.

We further discussed that, together with the lack of an example for an FIB Gaussian state, this leads us to conjecture that FIB Gaussian states do not exist, and thus all fully inseparable Gaussian states are, in fact, genuinely multipartite entangled. If correct, the conjecture would vastly simplify the detection of GME in Gaussian states, as the CM tests of full inseparability would be sufficient to detect GME. This would also have a practical impact on the experimental realization of multimode entanglement, as the experimental preparation of fully inseparable Gaussian states of a large number of modes is already possible~\cite{YokoyamaEtAl2013, ChenMenicucciPfister14, GerkeSperlingVogelCaiRoslundTrepsFabre15, YoshikawaYokoyamaKajiSornphiphatphongShiozawaMakinoFurusawa16}.

However, a rigorous proof of or a counterexample to the presented conjecture is still missing, and the question if FIB Gaussian states exist is thus left open for further investigation.


\begin{acknowledgments}
{\noindent}We thank G\'eza Giedke for useful discussions. 
J.P. acknowledges the use of the computational cluster of the Department of Optics. We acknowledge the use of several open-source software libraries \cite{mosek, HarrisEtAl2020, VirtanenEtAl2020, DalcinFang2021, SagnolStahlberg2022, RogowskiAseeriKeyesDalcin2023} in the computation and subsequent evaluation of the presented results.
K.B. and N.F. acknowledge support from the Austrian Science Fund (FWF) through the project P 36478-N funded by the European Union{\textemdash}NextGenerationEU.  
N.F. acknowledges support from the Austrian Federal Ministry of Education, Science and Research via the Austrian Research Promotion Agency (FFG) through the flagship project HPQC (FO999897481), and the projects FO999914030 (MUSIQ), FO999921407 (HDcode), and FO999921415 (Vanessa-QC), funded by the European Union{\textemdash}NextGenerationEU.
O.L. acknowledges support from Palack\'y University, a grant no. IGA-PrF-2026-005.
\end{acknowledgments}

\section*{Data and code availability}
The data and code that support the results of this study are available from the corresponding authors upon request.


\appendix

\section{Calculating density-matrix elements of a Gaussian state from its CM}\label{app-sec:DMmapping}

{\noindent}In this section, we discuss several methods that can be used to obtain density-matrix elements with respect to the Fock basis of a Gaussian state given by a CM~$\gamma$ and a vector of first moments~$\mathbf{d}$. The methods we discuss here are based on the quasiprobability distributions of the Gaussian states in question and the quasiprobability distributions of Fock states, and calculating the corresponding required density-matrix elements with respect to the Fock basis.\\

The most straightforward method is to start with the Wigner function of the Gaussian state~(\ref{eq:wignerGauss}) and find its overlap with the Wigner function of the corresponding Fock states by multiplying those Wigner functions and integrating over all phase-space variables. The sought-after density-matrix elements of a three-mode Gaussian state are then given by
\begin{align}\label{eq:overlap}
    \bra{k\subtiny{0}{0}{A}l\subtiny{0}{0}{B}m\subtiny{0}{0}{C}}&\rho\subtiny{0}{0}{ABC}^G\ket{k'\subtiny{0}{0}{A}l'\subtiny{0}{0}{B}m'\subtiny{0}{0}{C}}\\
    &=(2\pi)^3\int\!\!\dd^3\mathbf{x}\,\dd^3\mathbf{p}\ W_G(\mathbf{x},\mathbf{p})\times\nonumber\\
    &\quad\times W(\mathbf{x},\mathbf{p})\left[ \ketbra{k\subtiny{0}{0}{A}}{k'\subtiny{0}{0}{A}}\otimes\ketbra{l\subtiny{0}{0}{B}}{l'\subtiny{0}{0}{B}}\otimes\ketbra{m\subtiny{0}{0}{C}}{m'\subtiny{0}{0}{C}} \right],\nonumber
\end{align}
where $W(\mathbf{x},\mathbf{p})\left[M\right]$ is the Wigner function of the matrix element in the argument in square brackets. 

While this method is straightforward, evaluating the expression~(\ref{eq:overlap}) can be relatively slow and resource-consuming compared to the following two methods.\\

Another method presented in~\cite{QuesadaHeltIzaacArrazolaShahrokhshahiMyersSabapathy2019} uses the overlap of the Husimi~$Q$ function of the Gaussian state and the Glauber{\textendash}Sudarshan~$P$ function of an operator composed of multimode Fock-state matrix elements $\ketbra{\mathbf{k}}{\mathbf{k'}}$. The~$Q$ function of a Gaussian $N$-mode state is given by
\begin{equation}\label{app-eq:Q}
    Q_G(\mathbf{\alpha})=\frac{1}{\sqrt{\det(\pi\,\sigma\subtiny{-1}{0}{Q})}}\exp\left(-\tfrac{1}{2}(\mathbf{\alpha}-\mathbf{\beta})^\dagger\sigma\subtiny{-1}{0}{Q}^{-1}(\mathbf{\alpha}-\mathbf{\beta})\right),
\end{equation}
where $\sigma\subtiny{-1}{0}{Q}=\sigma+\tfrac{1}{2}\mathbb{1}$, and $\sigma$ is the covariance matrix with components $\sigma_{jk}=\Tr\left[\rho\{\zeta_j,\zeta_k\}\right]/2-\beta_j\beta_k$, with $\zeta=(a_A,a_B,a_C,a^\dagger_A,a^\dagger_B,a^\dagger_C)$ being a vector of annihialtion and creation operators and $\beta$ the vector of the first moments of $\zeta$. Since the annihilation and creation operators are related to the quadrature operators as $a_i=(x_i+ip_i)/\sqrt{2}$ and $a^\dagger_i=(x_i-ip_i)/\sqrt{2}$, one can obtain the three-mode CM $\sigma$ by way of a transformation of the CM $\gamma$ according to
\begin{equation}
    \sigma=\tfrac{1}{2}S\gamma S^\dagger
\end{equation}
with
\begin{equation}
    S=\frac{1}{\sqrt{2}}\begin{pmatrix}
        1 & 0 & 0 & i & 0 & 0 \\
        0 & 1 & 0 & 0 & i & 0 \\
        0 & 0 & 1 & 0 & 0 & i \\
        1 & 0 & 0 & -i & 0 & 0 \\
        0 & 1 & 0 & 0 & -i & 0 \\
        0 & 0 & 1 & 0 & 0 & -i
    \end{pmatrix}.
\end{equation}
The vector $\mathbf{\alpha}$ is given by $\mathbf{\alpha}^\top=(\alpha_A,\alpha_B,\alpha_C,\alpha^*_A,\alpha_B^*,\alpha_C^*)$, and $\mathbf{\beta}$ is a vector of mean values of the annihilation and creation operators. The elements of the density matrix are then given by the integral
\begin{equation}\label{app-eq:PQ}
    \brakket{\mathbf{k}}{\rho}{\mathbf{k'}}=\pi^3\int\!\!\dd\mathbf{\alpha}\,P_{\ketbra{\mathbf{k}}{\mathbf{k'}}}(\mathbf{\alpha})Q_G(\mathbf{\alpha}).
\end{equation}
This integral can be simplified if we define the symmetric matrix $A$ as
\begin{equation}\label{app-eq:AX}
    A=X(\mathbb{1}-\sigma\subtiny{-1}{0}{Q}^{-1}),\quad\text{with}\quad X=\mqty(\mathbb{0} & \mathbb{1} \\ \mathbb{1} & \mathbb{0})
\end{equation}
along with the vector $\mathbf{\vartheta}^\top=\mathbf{\beta}^\dagger\sigma\subtiny{-1}{0}{Q}^{-1}$. Then, the elements of the density matrix of the Gaussian state are given as
\begin{equation}\label{eq:QuesadaElem}
    \brakket{\mathbf{k}}{\rho}{\mathbf{k'}}=T\prod_{j=1}^3\partial_{\alpha_j}^{k_j}\partial_{\alpha^*_j}^{k'_j}\exp\eval{\left(\tfrac{1}{2}\mathbf{\alpha}^\top A\mathbf{\alpha}+\mathbf{\vartheta}^\top \mathbf{\alpha}\right)}_{\mathbf{\alpha}=0},
\end{equation}
with the prefactors collected into
\begin{equation}\label{app-eq:QuesadaPref}
    T=\frac{\exp\left(-\tfrac{1}{2}\mathbf{\beta}^\dagger\sigma\subtiny{-1}{0}{Q}^{-1}\mathbf{\beta}\right)}{\sqrt{\det(\sigma\subtiny{-1}{0}{Q})\prod_{j=1}^3k_j!\,k'_j!}}.
\end{equation}
More details of the derivation of Eq.~(\ref{eq:QuesadaElem}) can be found in~\cite{QuesadaHeltIzaacArrazolaShahrokhshahiMyersSabapathy2019}. The evaluation of Eq.~(\ref{eq:QuesadaElem}) is much faster than the previous method, but it still becomes relatively time and resource-consuming with increasing excitation number of the Fock states required to construct the desired density-matrix elements.\\

The last and for us the most efficient method derives the overlap of Wigner functions~(\ref{eq:overlap}) in terms of multidimensional Hermite polynomials~\cite{DodonovMankoManko1994}. The multidimensional Hermite polynomial is defined by a vector $\mathbf{y}$ dependent on the CM $\gamma'$ and the vector of first moments $\mathbf{d}'$ of the Gaussian state as
\begin{equation}\label{app-eq:y}
    \mathbf{y}=U(\mathbb{1}+2\gamma')^{-1}\mathbf{d'},
\end{equation}
where both the CM $\gamma'$ and the vector of first moments $\mathbf{d}'$ are taken to be in an $ppxx$-ordering and
\begin{equation}\label{app-eq:U}
    U=\left(\begin{array}{cc}
        -i\mathbb{1} & \mathbb{1} \\
        \phantom{-}i\mathbb{1} & \mathbb{1}
    \end{array}\right).
\end{equation}
In addition to the vector $\mathbf{y}$, the definition of the multidimensional Hermite polynomial also relies on the definition of a symmetric matrix~$r$ given by
\begin{equation}\label{app-eq:R}
    R=U^\dagger(\mathbb{1}-2\gamma')(\mathbb{1}+2\gamma')^{-1}U^*.
\end{equation}
The elements of the density matrix of the state are then given by
\begin{equation}\label{eq:Hermite}
    \brakket{\mathbf{k}}{\rho}{\mathbf{k'}}=\frac{2^NH_{\mathbf{k},\mathbf{k'}}^R(\mathbf{y})}{\sqrt{\det(\gamma'+\mathbb{1})\prod_{j=1}^Nk_j!\,k'_j!}}.
\end{equation}
In our implementation, the evaluation of Eq.~(\ref{eq:Hermite}) is the fastest among the three presented ways and therefore it appears to be the most suitable for calculations of larger parts of the density matrix of the Gaussian state. Thus, this is the method we used for our calculations, with double-checks using Eq.~(\ref{eq:QuesadaElem}).


\bibliographystyle{apsrev4-1fixed_with_article_titles_full_names_new}
\bibliography{Master_Bib_File}


\end{document}